\documentclass[%
 reprint,
 amsmath,amssymb,
 aps,
]{revtex4-2}

\usepackage{graphicx}% Include figure files
\usepackage{dcolumn}% Align table columns on decimal point
\usepackage{bm}% bold math
\begin{document}

\preprint{APS/123-QED}

\title{High-order mesh-free direct numerical simulation of lean hydrogen flames in confined geometries}% Force line breaks with \\

\author{H. M. Broadley}
\email{henry.broadley@manchester.ac.uk}
\author{J. R. C. King}%
\affiliation{%
 School of Engineering, The University of Manchester}%


\author{S. J. Lind}
\affiliation{School of Engineering, Cardiff University}%

\date{\today}% It is always \today, today,

\begin{abstract}
Here we perform the first analysis of high-fidelity simulations of the propagation of lean hydrogen flames through porous media, taking cylindrical arrays a representative example geometry. In this fundamental study we discuss the impact of confinement on both thermodiffusive and thermoacoustic instabilities. Flame propagation in these complex geometries is cannot be performed by leading mesh-based codes, and is instead simulated using a high-order meshfree method, LABFM.  Pore scale propagation is shown to be dependent on throat width between cylinders, and this is then related to large-scale flame dynamics, allowing us to give a heuristic explanation for the increased growth rate of the thermodiffusive instability in more confined geometries. Thermoacoustic instabilities are also observed for sufficiently confined geometries. Understanding these instability mechanisms is crucial for improving the design of future combustors, both in terms of controlling flame dynamics and increasing the durability of combustors.
%\begin{description}
%use the optional argument of the \verb+\item+ command to %give the category of each item. 
%\end{description}
\end{abstract}

\maketitle

%\tableofcontents

\section{Introduction}\addvspace{10pt}

%% Previous DNS.
The highly multiscale nature of turbulent combustion, where sub-millimetre flame fronts interact non-linearly with large scale hydrodynamic, turbulent and acoustic structures, poses significant challenges for direct numerical simulation (DNS). Any DNS method needs to be both highly accurate and computationally efficient. High-order methods are widely accepted as necessary~\cite{emmett_2019,domingo_2023}, providing equivalent accuracy at lower resolution (and hence cost) than their low-order counterparts. For an overview of the state-of-the-art in combustion DNS, we direct the reader to~\cite{domingo_2023}. The requirements of high accuracy and computational efficiency have led the field to be dominated by high-order finite difference codes, such as SENGA+~\cite{cant_2012}, S3D~\cite{chen_2011}, and KARFS~\cite{perez_2018}, based on tenth- and eighth-order central finite differences, with explicit Runge-Kutta schemes for time integration.

Finite-difference codes such as these are by nature limited to simple geometries, although non-trivial geometries may be included at low-order via immersed boundary methods~\cite{rauch_2018}. Given the drive for geometric flexibility in the next generation of combustion codes, more flexible alternatives have been sought. Unstructured mesh-methods provide increased geometric flexibility (e.g. the code AVBP~\cite{migbreb_2016}), although they are often limited to low order spatial discretisations. In recent years the spectral difference method~\cite{marchal_2023} has emerged as a promising candidate to deliver high-order, adaptive simulations on unstructured meshes in real geometries~\cite{domingo_2023}. Whilst the need to capture acoustic waves renders weak form methods (in which a large sparse linear system must be solved every time step) unfeasible, in the arena of low-Mach solvers, weak form methods can be beneficial. A prime example here is the leading Spectral Element method (SEM) code Nek5000~\cite{nek5000}, which has been applied to low-Mach DNS of combustion in complex geometries (e.g.~\cite{schmitt_2016,danciu_2025}), with exceptionally high accuracy. A limitation common to all unstructured-mesh methods is the challenge of constructing a high-quality body-fitted mesh. In highly complex geometries, creating a high-quality mesh, without significant deterioration of accuracy due to skewed mesh elements, can be extremely resource intensive, in some cases (e.g. porous media), taking longer than the simulation itself~\cite{wood_2020}. Studies of methane flames in porous media have been performed using finite-volume methods (see \cite{ferguson_2021}, \cite{boigne_2024}) which are typically second-order accurate. These have focused on the impact of conjugate heat transfer and hydrodynamic effects, yielding insight into the changes in flame dynamics due to pore-flame interactions. High-order methods capable of simulating flows in these geometries will support the exploration of these rich dynamics. 

% Mesh-free methods
We approach the problem from a fundamentally different perspective: mesh-free methods. Mesh-free methods, in which no information on the topological connectivity of computational points (also/often referred to as particles) is required, are an alternative to grid-based or mesh-based methods. Whilst Smoothed Particle Hydrodynamics (SPH) is the most widely used (see~\cite{lind_2020} for a recent review), there are many variations, including Radial Basis Function (RBF) methods, generalised finite difference methods (GFDM), generalised moving least squares (GMLS) methods, the Reproducing Kernel Particle Method (RKPM), and Finite Particle Methods (FPM). For an overview of these, we refer the reader to our previous work~\cite{king_2020}. With no information on the topological connectivity between collocation points, discretising a complex geometry with an unstructured node-set (e.g. using propagating front algorithms~\cite{fornberg_2015a}) is straightforward and can be easily automated. A limitation of the majority of the above methods is accuracy, with most restricted by construction to low-order convergence with increasing resolution. 

To address this, the Local Anisotropic Basis Function method (LABFM)~\cite{king_2020}, a high-order mesh-free method recently developed by some of the authors, is used to approximate spatial derivatives. LABFM is a method for obtaining finite-difference style weights on a stencil of arbitrarily distributed unstructured nodes, with no information on node connectivity required. The weights are obtained by solution of local linear systems constructed to ensure that polynomial consistency is satisfied up to the desired order. Compared with other consistent mesh-free methods (for a review of which, we refer the reader to the introductions of~\cite{king_2020,king_2022}), LABFM may be formulated at higher order, and has a lower computational cost for a given level of accuracy. The combination of scalability, geometric flexibility, and high accuracy, render LABFM a promising method for the next generation of combustion DNS codes. A LABFM-based framework for combustion DNS was introduced in~\cite{king_2024}.

% What we show here
In this work, we use the LABFM framework for combustion DNS to investigate the dynamics of lean hydrogen flames in confined geometries. The implementation is termed the SUNSET-code (\textbf{S}calable \textbf{U}nstructured \textbf{N}ode-\textbf{SET} code), and has been extensively validated for a range of two- and three-dimensional cases and on both local and national high-performance computing (HPC) systems, exhibiting both excellent accuracy and scaling. The SUNSET code is available open-source at~\cite{sunset}. Next we introduce the governing equations solved, followed by a description of the numerical method used. We then present the results of a numerical investigation into lean hydrogen flame behaviour in confined geometries using this numerical framework, followed by a summary of conclusions.

\section{Governing Equations}\label{sec:ge}

We numerically solve the compressible Navier-Stokes equations, which may be expressed in conservative form for a mixture of $N_{S}$ miscible reacting species of semi-perfect gases as
\begin{subequations}
\begin{align}
\frac{\partial\rho}{\partial{t}}+\frac{\partial\rho{u}_{k}}{\partial{x}_{k}}=&~0\label{eq:mass}\\
\frac{\partial\rho{u}_{i}}{\partial{t}}+\frac{\partial\rho{u}_{i}u_{k}}{\partial{x}_{k}}=&-\frac{\partial{p}}{\partial{x}_{i}}+\frac{\partial\tau_{ki}}{\partial{x}_{k}}\label{eq:mom}\\
\frac{\partial\rho{E}}{\partial{t}}+\frac{\partial\rho{u}_{k}E}{\partial{x}_{k}}=&-\frac{\partial{p}u_{k}}{\partial{x}_{k}}-\frac{\partial{q}_{k}}{\partial{x}_{k}}+\frac{\partial\tau_{ki}u_{i}}{\partial{x}_{k}}\label{eq:en}\\
\frac{\partial\rho{Y}_{\alpha}}{\partial{t}}+\frac{\partial\rho{u}_{k}Y_{\alpha}}{\partial{x}_{k}}=&~\dot\omega_{\alpha}-\frac{\partial\rho{V}_{\alpha,k}Y_{\alpha}}{\partial{x}_{k}}\label{eq:Y}\\& \qquad \qquad \qquad\forall\alpha\in\left[1,N_{S}\right]\notag
\end{align}\label{eq:ge}
\end{subequations}
where $\rho$ is the density, ${u}_{i}$ the $i$-th component of velocity, $p$ is the pressure, $E$ is the total energy, $\tau_{ki}$ are components of the viscous stress tensor, $q_{k}$ the $k$-th component of the heat flux vector, $Y_{\alpha}$ is the mass fraction of species $\alpha\in\left[1,N_{S}\right]$, and $V_{\alpha,k}$ is the $k$-th component of the molecular diffusion velocity of species $\alpha$. $\dot\omega_{\alpha}$ is the production rate of species $\alpha$. The total energy is related to the other thermodynamic quantities by
\begin{equation}E=\displaystyle\sum_{\alpha}h_{\alpha}Y_{\alpha}-\frac{p}{\rho}+\frac{1}{2}{u}_{i}{u}_{i},\label{eq:E}\end{equation}
where $h_{\alpha}$ is the enthalpy of species $\alpha$. In~\eqref{eq:E} and hereafter, sums over $\alpha$ are taken to be over all species $\alpha\in\left[1,N_{S}\right]$. 
%The viscous stress is defined as 
%\begin{equation}\tau_{ki}=\mu\left(\frac{\partial{u}_{k}}{\partial{x}_{i}}+\frac{\partial{u}_{i}}{\partial{x}_{k}}-\frac{2}{3}\frac{\partial{u}_{j}}{\partial{x}_{j}}\delta_{ik}\right)\label{eq:tau},\end{equation}
%where $\mu$ is the dynamic viscosity and $\delta_{ik}$ is the Kronecker delta. The heat flux vector is given by
%\begin{equation}q_{k}=-\lambda\frac{\partial{T}}{\partial{x}_{k}}+\displaystyle\sum_{\alpha}\rho{V}_{\alpha,k}Y_{\alpha}h_{\alpha}\label{eq:hfv},\end{equation}
%where $\lambda$ is the thermal conductivity of the mixture.
The temperature dependencies of thermodynamic quantities (heat capacity $c_{p,\alpha}$,  enthalpy $h_{\alpha}$) for each species take polynomial form fitting the standard NASA polynomials~\cite{gordon_1971}. We use a mixture-averaged model for transport properties, following combination rules of~\cite{ern_1994,ern_1995}, with the Hirschfelder-Curtiss approximation for molecular diffusion~\cite{hirschfelder_curtiss}. Soret and Dufour effects are neglected as in \cite{Howarth_2022}. In this work we focus on hydrogen flames, using the $21$ step, $9$ species hydrogen-air reaction mechanism of~\cite{li_2004}. As an aside we note that the single-step irreversible Arrhenius mechanism of \cite{dominguez_2023}, constructed for dilute hydrogen combustion, was used to simulate combustion flows in the same geometries of $\S4$, however these flames were observed to be prone to thermoacoustic instabilities. This behaviour will be discussed in a future study.

\section{Numerical Methods}\label{sec:nm}

The numerical implementation follows that described in~\cite{king_2024}, to which we refer the reader for complete details. The domain is discretised with an unstructured node-set, and spatial derivatives are calculated using the Local Anisotropic Basis Function Method (LABFM). Boundary conditions are imposed using the Navier-Stokes characteristic boundary condition formalism (following~\cite{sutherland_2003,yoo_2007}), with one-sided difference operators at non-periodic boundaries. Time integration is by explicit third order Runge-Kutta scheme. In common with other high-order collocated methods, the discretisation admits solutions with energy at the wavenumber of the resolution, and we de-alias the solution at each time step by applying a high-order low-pass filter to the conservative variables. 

\subsection{Spatial discretisation}

For details and extensive analysis of LABFM we refer the reader to ~\cite{king_2020,king_2022}. Here, we provide the minimum detail sufficient for reproduction of the method. We consider only two-dimensional computational domains, which are discretised with a point cloud of $N$ nodes. This point cloud is unstructured internally, and has local structure near boundaries. The local node spacing can vary along walls, inflow and outflow boundaries, near which an additional $4$ rows of uniformly distributed nodes are arranged along boundary normals originating at the boundary nodes. The internal point cloud is generated using the propagating front algorithm of~{\cite{fornberg_2015a}}. Each node $a$ has an associated resolution $s_{a}$, corresponding to the local average distance between nodes. The resolution need not be uniform. Each node also has an associated computational stencil length-scale $h_{a}$, which again may vary spatially. The ratio $s_{a}/h_{a}$ is approximately uniform, with some variation due to the stencil optimisation procedure described in~\cite{king_2022}, which ensures the method uses stencils marginally larger than the smallest stable stencil. Each node holds the evolved variables $\rho_{a}$, $\left(\rho{u}_{i}\right)_{a}$, $\left(\rho{E}\right)_{a}$, and $\left(\rho{Y}_{\alpha}\right)_{a}$. The governing equations are solved on the set of $N$ nodes. 

\begin{figure}
    \centering
    \includegraphics[width=0.44\textwidth]{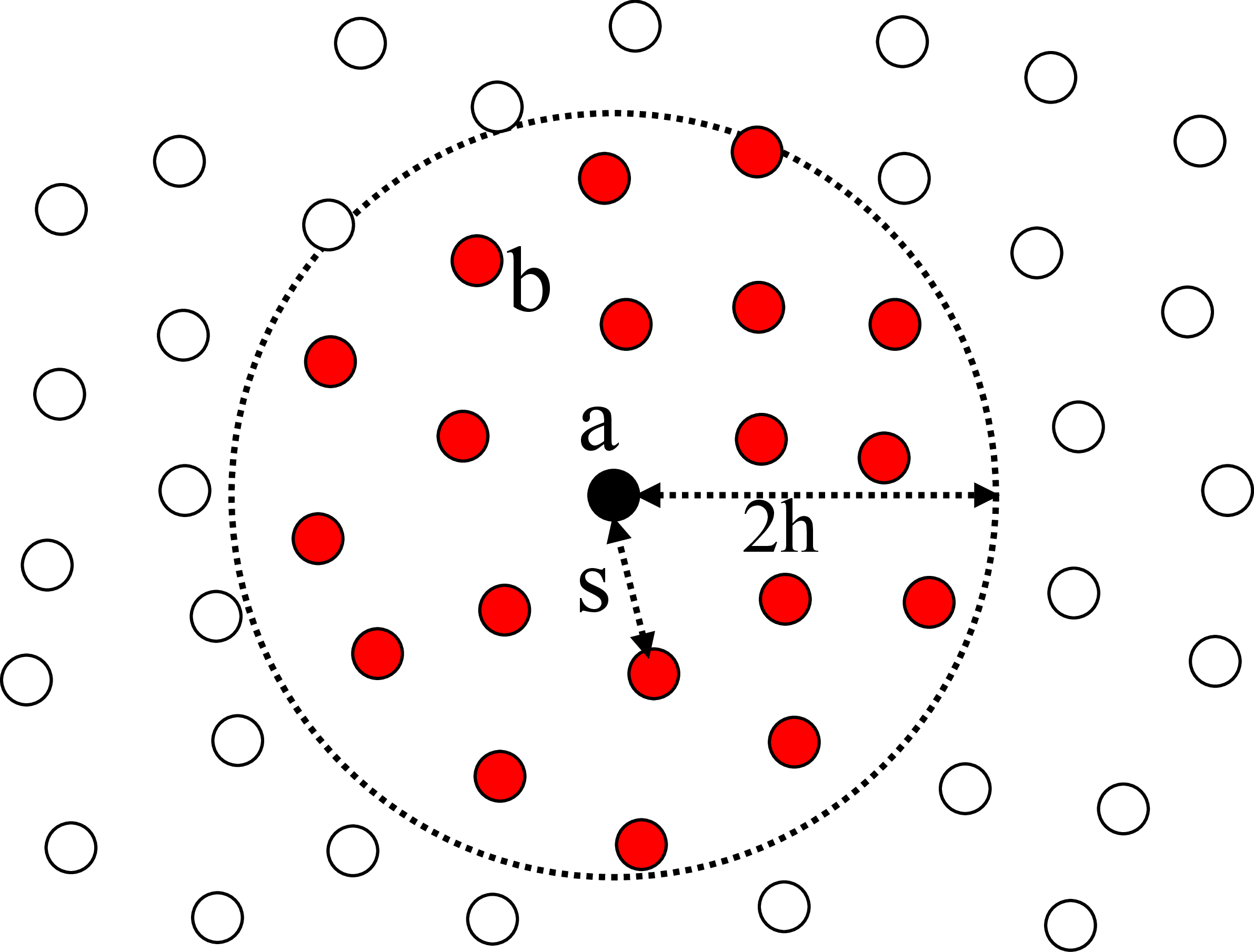}
    \caption{A schematic of the computational stencil.\label{fig:stencil}}    
\end{figure}

The difference between properties at two nodes is denoted $\left(\cdot\right)_{ba}=\left(\cdot\right)_{b}-\left(\cdot\right)_{a}$. The computational stencil for each node $a$ is denoted $\mathcal{N}_{a}$, and contains all nodes $b$ for which {$r^{2}_{ba}={x}_{ba}^{2}+y_{ba}^{2}\le4h_{a}^{2}$}. A schematic of the stencil is shown in Fig.~\ref{fig:stencil}. In the following, we use standard multi-index notation with index (e.g.) $\alpha$ representing the ordered pair $\left(\alpha_{1},\alpha_{2}\right)$, with $\left\lvert{\alpha}\right\rvert=\alpha_{1}+\alpha_{2}$.
All spatial derivative operators take the form
\begin{equation}L^{\gamma}_{a}\left(\cdot\right)=\displaystyle\sum_{b\in\mathcal{N}_{a}}\left(\cdot\right)_{ba}w^{\gamma}_{ba},\label{eq:general_do}\end{equation}
where $\gamma$ is a multi-index identifying the derivative being approximated by~\eqref{eq:general_do}, and $w^{\gamma}_{ba}$ are a set of inter-node weights. This construction describes both LABFM and high-order finite difference operators. To evaluate $w^{\gamma}_{ba}$ for LABFM operators, we first define element-wise a vector of monomials $\bm{X}^{\alpha}_{ba}$ and a vector of anisotropic basis functions $\bm{W}^{\alpha}_{ba}$, with the elements corresponding to multi-index $\alpha$ given by
\begin{equation}\bm{X}^{\alpha}_{ba}=\frac{x_{ba}^{\alpha}}{\alpha!},\quad\bm{W}^{\alpha}_{ba}=\frac{\psi\left({r}_{ba}/h_{a}\right)}{\sqrt{2^{\left\lvert{\alpha}\right\rvert}}}H_{\alpha}\left(\frac{x_{ba}}{h_{a}\sqrt{2}}\right),\end{equation}
where $H_{\alpha}$ are bi-variate Hermite polynomials of the physicists kind, and the radial basis function (RBF) $\psi$ is a Wendland C2 kernel~\cite{dehnen_aly}. The weights $w_{ba}^{\gamma}$ in~\eqref{eq:general_do} are constructed as
\begin{equation}w^{\gamma}_{ba}=\bm{W}_{ba}\cdot\bm{\Psi}^{\gamma},\end{equation}
with $\bm{\Psi}^{\gamma}$ a vector to be determined by solution of the system
\begin{equation}\left[\displaystyle\sum_{b\in\mathcal{N}_{a}}\bm{X}_{ba}\otimes\bm{W}_{ba}\right]\cdot\bm{\Psi}^{\gamma}=\bm{C}_{\gamma},\label{eq:lsys}\end{equation}
in which $\bm{C}_{\gamma}$ is a unit vector defined element-wise as $C_{\gamma}^{\alpha}=\delta_{\alpha\gamma}$. 

The consistency of the operator~\eqref{eq:general_do} is determined by the size of the linear system~\eqref{eq:lsys}. If we desire polynomial consistency of order $m$, we include all terms with $\lvert\alpha\rvert\le{m}$. For each node we construct and solve the linear system~\eqref{eq:lsys} to obtain $\bm{\Psi}^{\gamma}$, for $\gamma$ corresponding to both first spatial derivatives, and the Laplacian, which we then use to calculate and store $w_{ba}^{\gamma}$ in~\eqref{eq:general_do}. %The right panel of Figure~\ref{fig:stencil} shows the convergence properties of the LABFM gradient operator. 
In the present work we set $m=8$ internally, dropping to $m=4$ at non-periodic boundaries.

\section{Hydrogen flames in a porous geometry}
\label{sec:res}

We focus on an idealised porous media, represented by a hexagonal array of cylinders. We present results for exclusively two-dimensional simulations, noting that the method has been validated for three-dimensional turbulent reacting flows in~\cite{king_2024}, and the effect of confinement for these three-dimensional flames is an active area of investigation. We consider a hydrogen-air mixture with an equivalence ratio of $\phi=0.32$, at atmospheric pressure, and a temperature of $300K$. The laminar flame speed is $S_{L}\approx0.07m/s$, and thermal flame thickness is $\delta_{L}\approx1$mm, which we use to non-dimensionalise the problem. We fix the diameter $D/\delta_L=5$ of each cylinder, choosing to vary the cylinder spacing $S/D$. The computational domain is rectangular with length $25-30D$ (depending on cylinder spacing) and width $10D$, with periodicity laterally, and non-reflecting inflow and outflow boundary conditions in the streamwise direction. On the cylinder surfaces we impose adiabatic no-slip conditions following~\cite{boigne_2024}, allowing us to isolate the hydrodynamic and thermoacoustic behaviour from instabilities related to conjugate heat transfer~\cite{yakovlev_2021}. The simulations are initialised with a laminar flame just downstream of the cylinder array, and subject to a small, random, multi-mode perturbation with amplitude $\delta_{L}/4$. We consider the effect of different cylinder spacings on flame dynamics. % A typical snapshot of the flame for the case with $S/D=5/2$ is shown in Figure \ref{flame_snapshot_example}.

%\begin{figure}
%    \centering
%    \includegraphics[width=0.49\textwidth]{alpha_SD52_t3.png}
%    \caption{Snapshot of the normalised heat release rate showing the location of the flame for confinement $S/D=5/2$ at $t^*\approx 10.5$.}
%   \label{flame_snapshot_example}
%\end{figure}

%\subsection{Early time behaviour of the flame}
%Although we are primarily interested in the flame dynamics as it propagates through the array, we briefly investigate the behaviour of the flame prior to it entering the array. Fig.\ref{Early_time_flames} shows snapshots of the heat release rate for each confinement prior to the flame entering the array. In the early time regime, increased confinement causes flames to be thicker, and the most confined flame is substantially cooler than in the other two cases. In addition, the flame in the most confined geometry attaches to the cylinder surfaces, whereas the flame is extinguished behind the cylinders in the $S/D=5/2,5/3$ cases.

%\begin{figure*}[h!]
%\centering
%\vspace{-0.4 in}
%\includegraphics[width=0.32\textwidth]{alpha_SD_52_ex.png}
%\includegraphics[width=0.32\textwidth]{alpha_SD_53_ex.png}
%\vspace{2 pt}
%\includegraphics[width=0.32\textwidth, height=2.45cm]{alpha_SD_54_ex.png}
%\caption{Snapshots of heat release rate showing flame behaviour prior to it entering the arrays. Left panel $S/D=5/2$, middle panel $S/D=5/3$, right panel $S/D=5/4$.}
%\label{Early_time_flames}
%\end{figure*}

\subsection{Flame Propagation Through Pores}
\label{sub_sec:pore-by-pore_propagations}
We first consider how flames propagate through pores of different sizes. We allow the flame to become established within a few rows of the porous matrix before focussing in on a specific region to observe the typical features of flame propagation through two parallel pores for a given value of $S/D$. In all cases we study how the flame evolves after a given dimensionless time $t^*=t^*_0$.

\begin{figure*}
\centering
%\vspace{-0.4 in}
\includegraphics[width=0.24\textwidth]{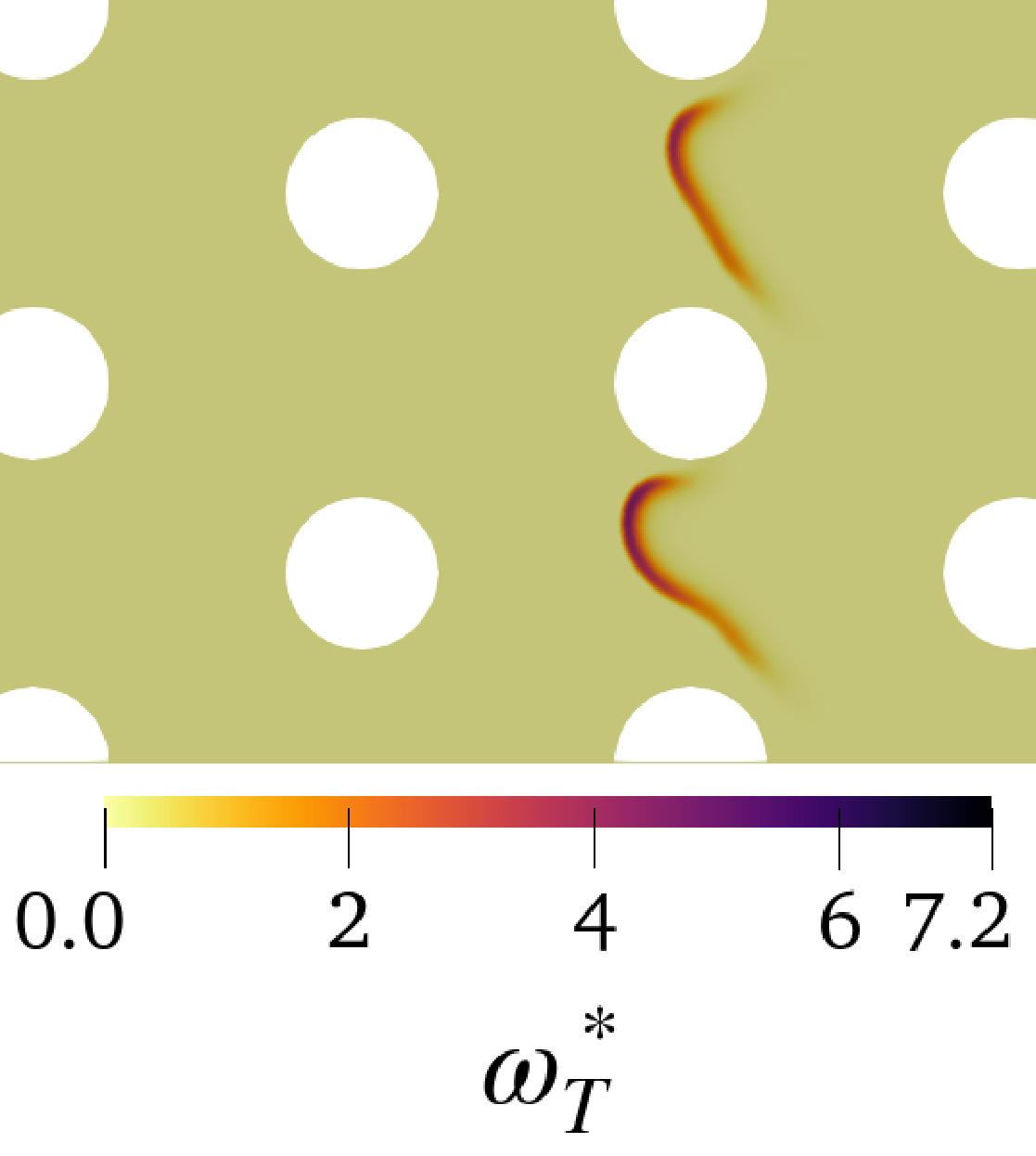}
\includegraphics[width=0.24\textwidth]{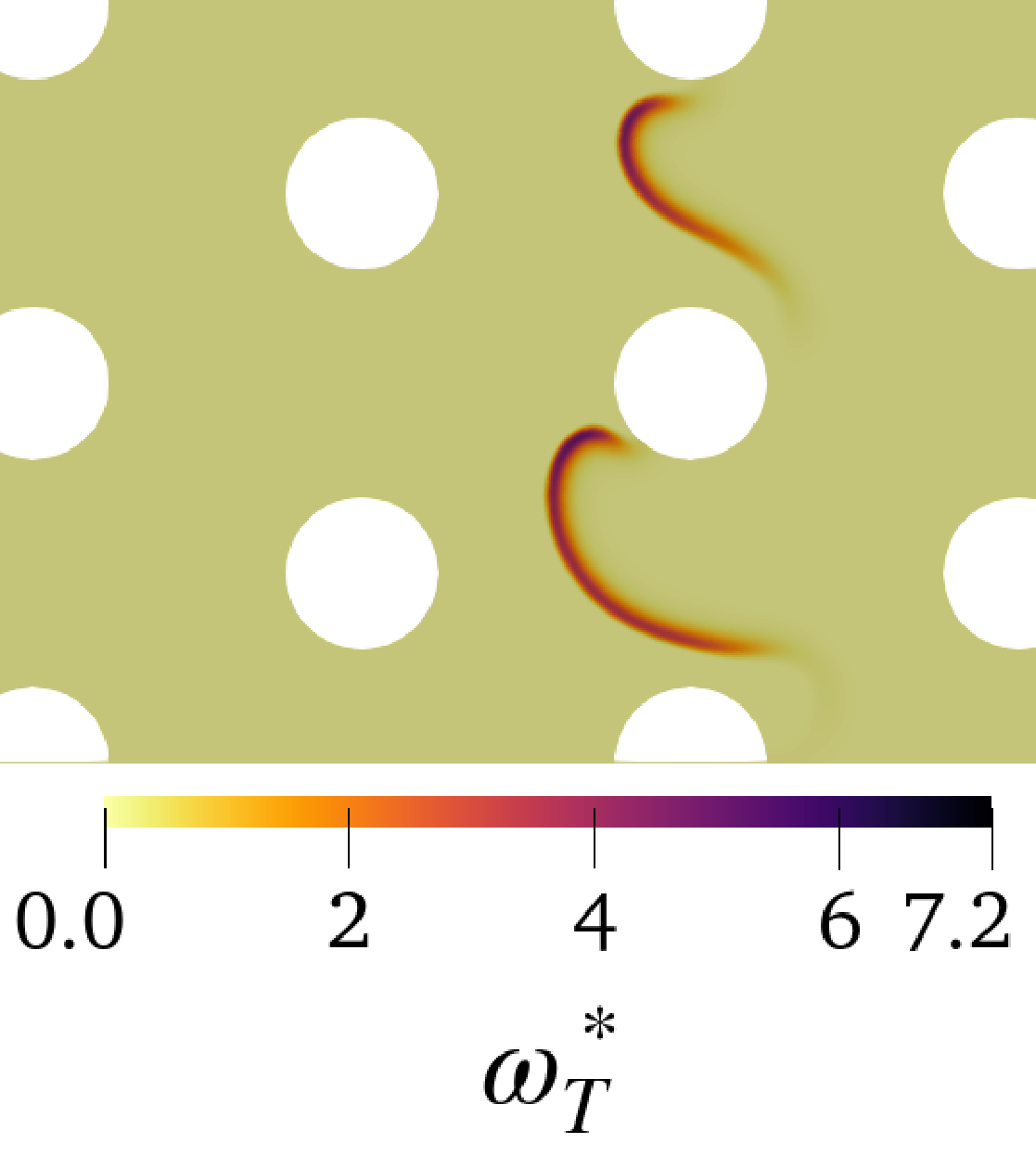}
\vspace{2 pt}
\includegraphics[width=0.24\textwidth]{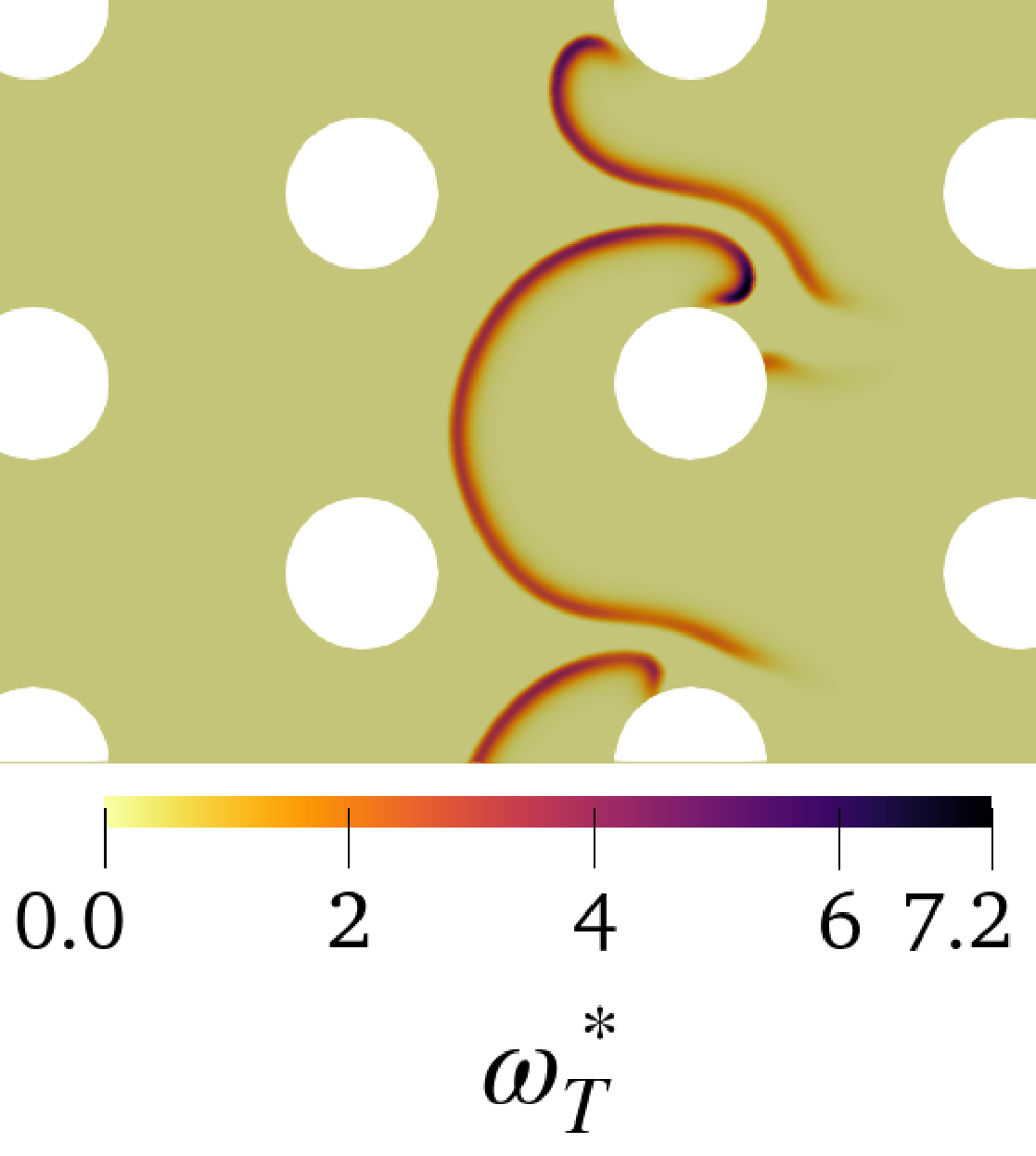}
\includegraphics[width=0.24\textwidth]{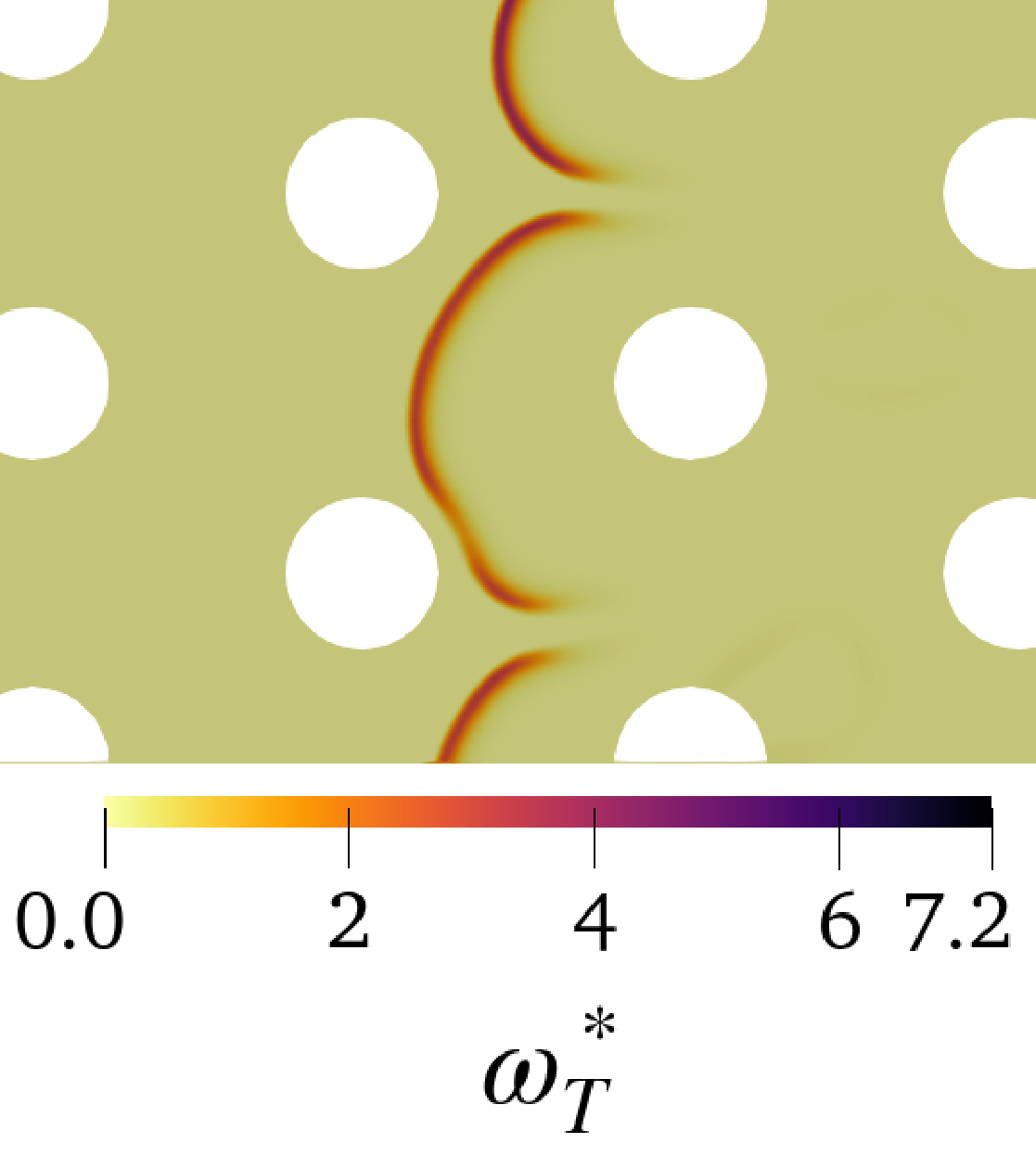}
\caption{Snapshots of heat release rate showing propagation of flame through typical pore with confinement $S/D=5/2$. From left to right: $t^*-t^*_0=0$, $t^*-t^*_0\approx0.56$, $t^*-t^*_0\approx1.33$, $t^*-t^*_0\approx1.75$.}
\label{fig:SD_52_pbp}
\end{figure*}

We begin by considering the least confined case, $S/D=5/2$. Fig.\ref{fig:SD_52_pbp} shows snapshots of heat release rate at four different times. Initially, the flame is propagating relatively evenly through each pore, though there is local extinction close to cylinder surfaces. However, as the flame continues to move through this part of the matrix we observe that in each pore the flame preferentially propagates on one side of the pore (near the upper cylinder), and in the bottom half of each pore there is a clear extinction region. As the flame evolves the preferentially propagating part burns quickly through the region where the flame was previously quenched (thus where a greater proportion of fuel likely is), with the maximum heat release rates attained here. The flame then appears to become approximately uniform as it moves towards the next row of cylinders.

\begin{figure*}
\centering
%\vspace{0.4 i}
\includegraphics[width=0.48\textwidth]{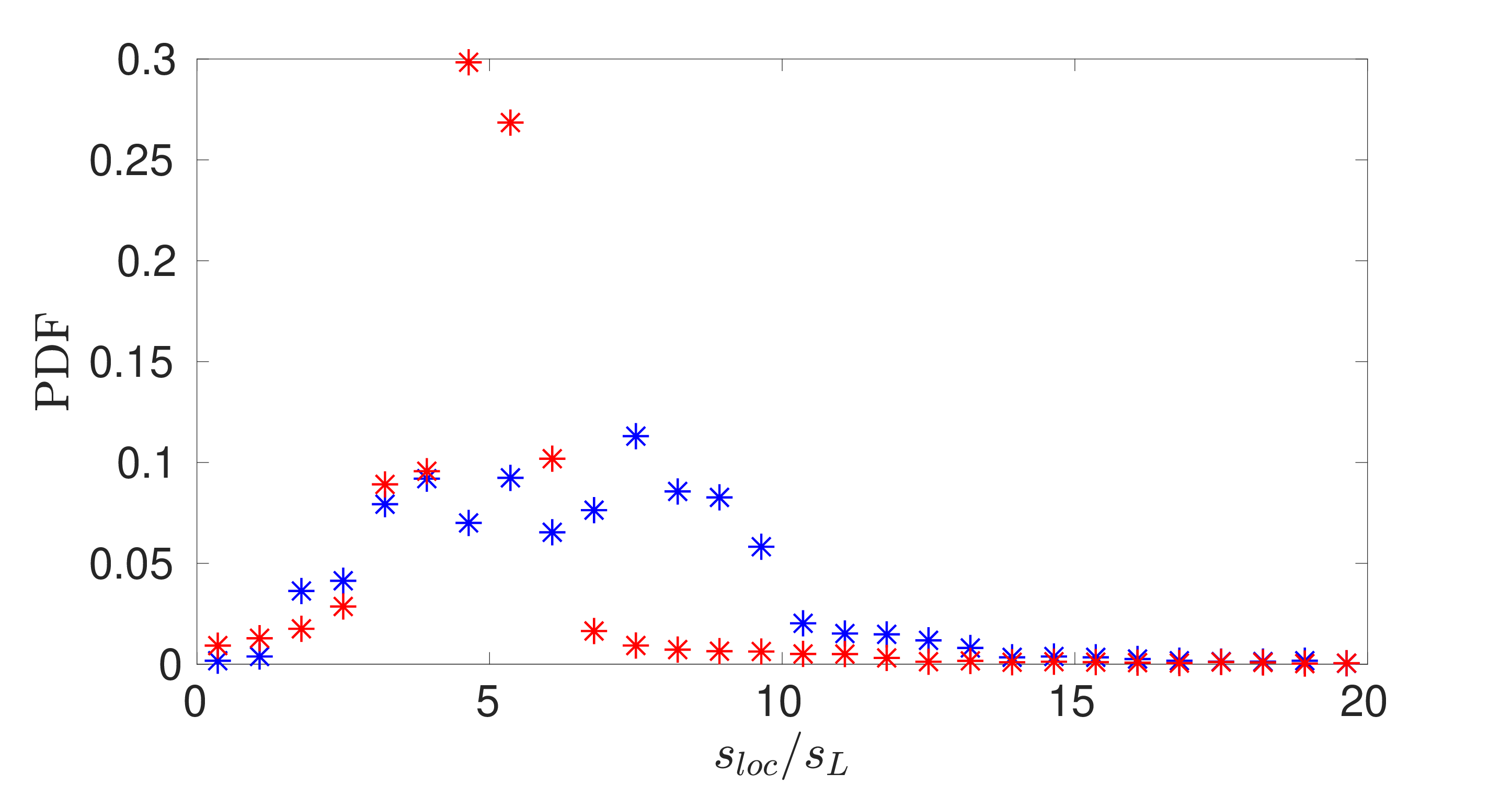}
\includegraphics[width=0.48\textwidth]{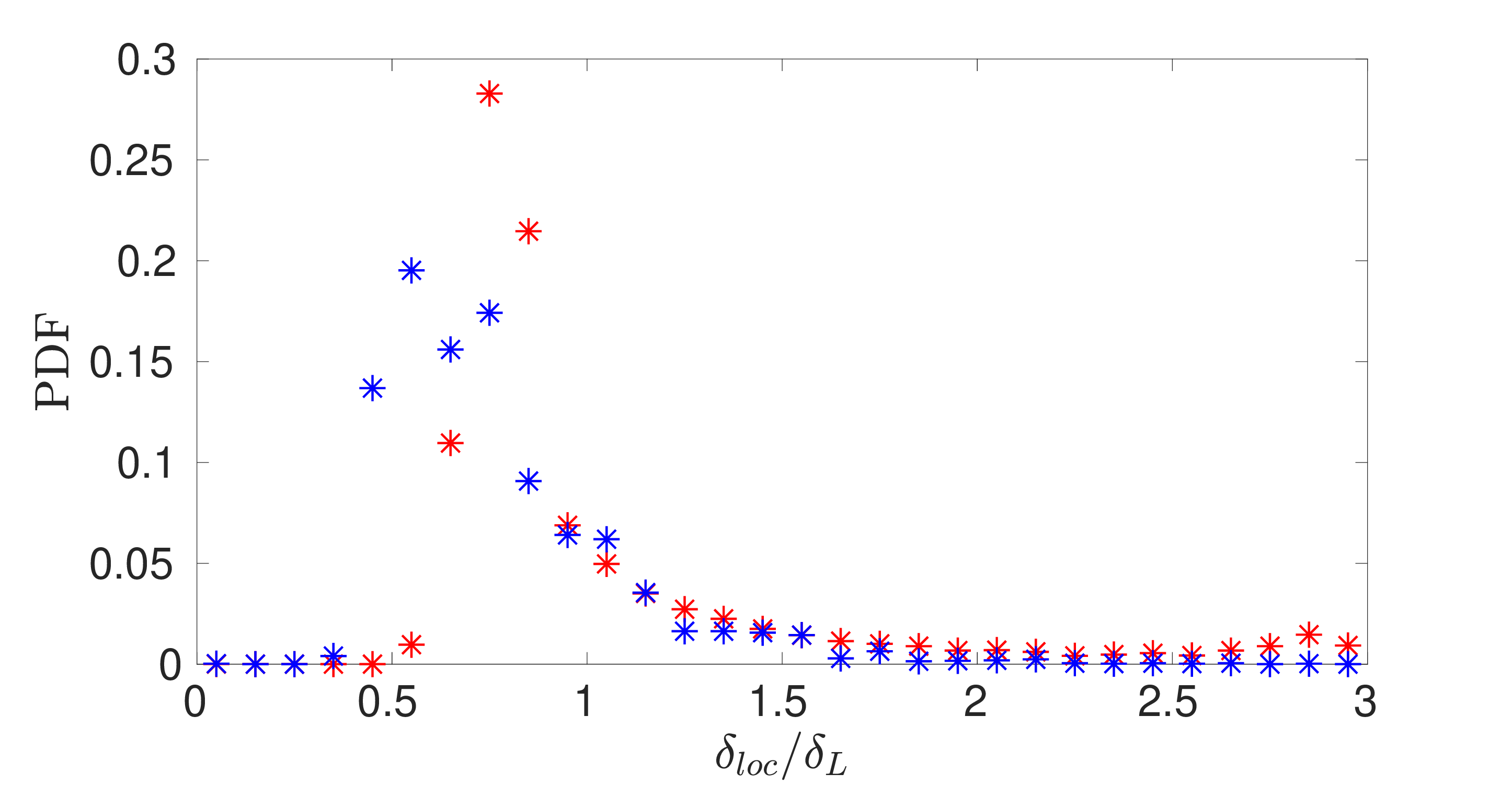}
\caption{PDF of normalised flame speed (left panel) and normalised flame thickness (right panel) for the part of the flame in the domain shown in Fig\ref{fig:SD_52_pbp} ($S/D=5/2$) for time period $0\leq t^*-t^*_0\leq 1.75$ (red $*$), and flame in domain shown in Fig\ref{fig:SD_54_pbp} ($S/D=5/4$) for time period $0\leq t^*-t_0^*\leq 0.98$ (blue $*$).}
\label{fig:flame_histograms}
\end{figure*}

It is also of interest to consider the local flame statistics in this region as the flame propagates through the pores. Defining the local flame speed $s_{loc}$ and local flame thickness $\delta_{loc}$ as in \cite{day_2009} with the progress variable as defined in \cite{Howarth_2022}, in Fig.\ref{fig:flame_histograms} we observe that over the time interval shown in Fig.\ref{fig:SD_52_pbp} the flame speed is considerably larger than the laminar value, and the flame is, on average, slightly thinner. These ensemble averages of statistics showed little dependence on time, suggesting that flame properties remain constant as the flame travels through pores of this size.

\begin{figure*}
\centering
%\vspace{0.4 i}
\includegraphics[width=0.24\textwidth]{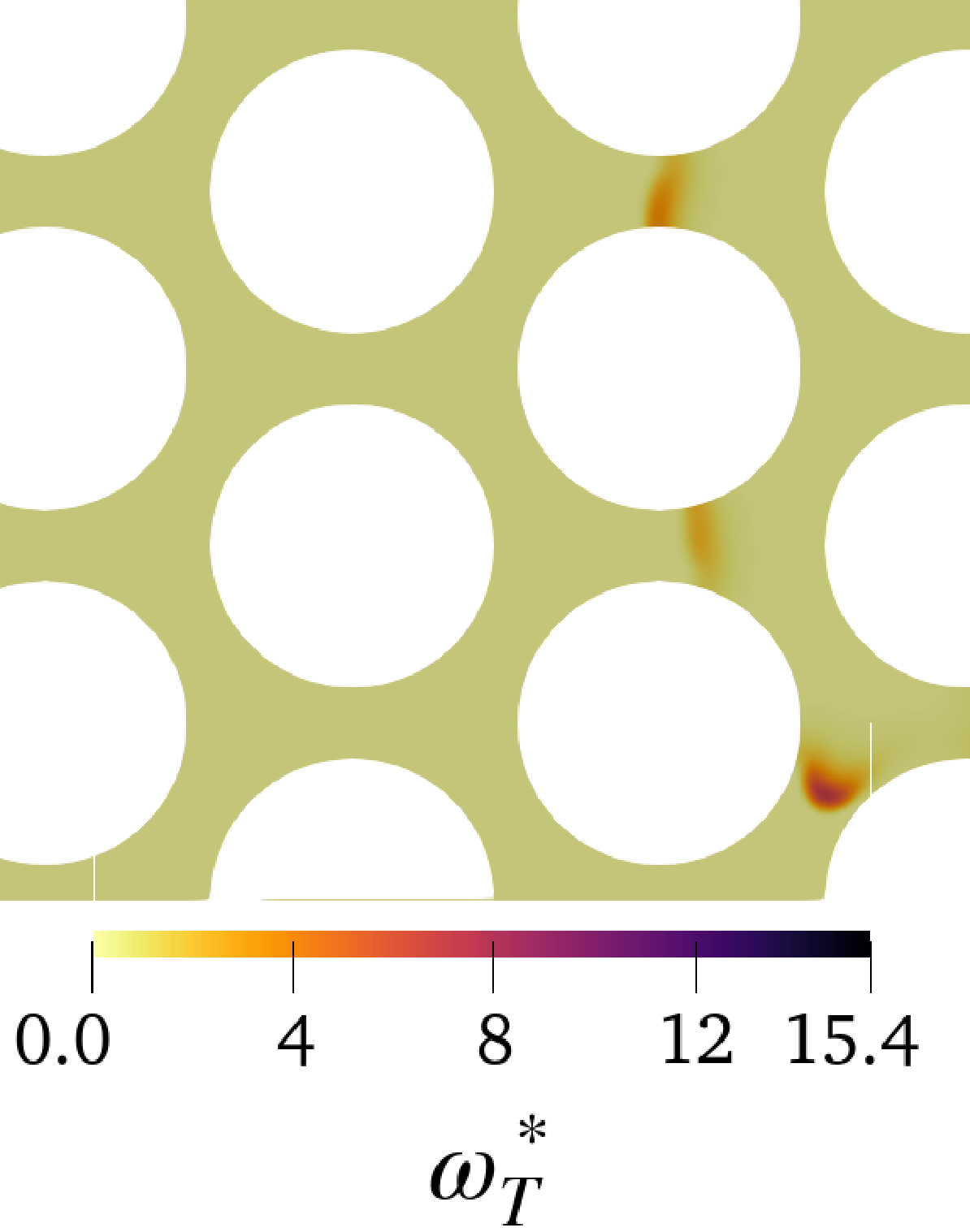}
\includegraphics[width=0.24\textwidth]{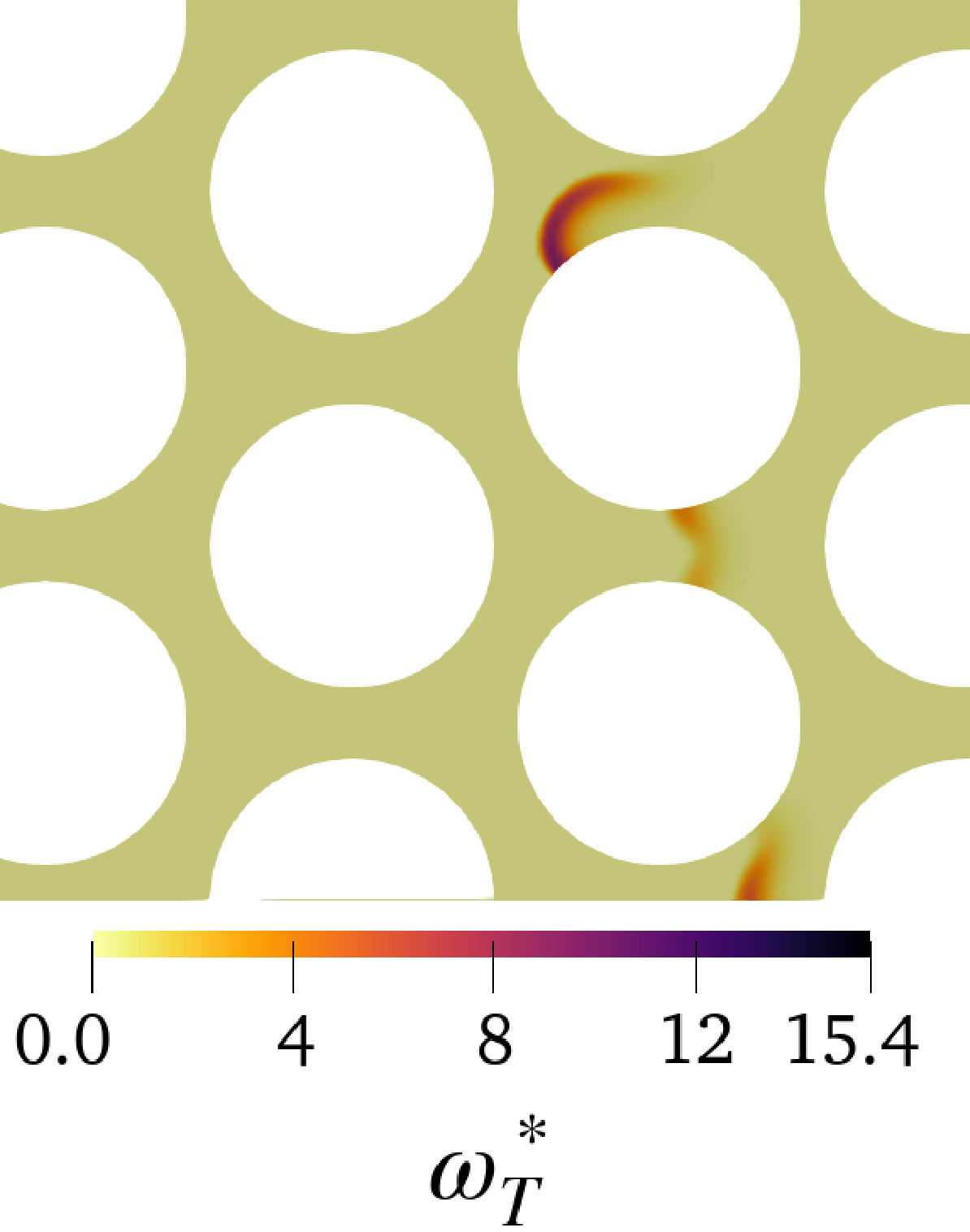}
\vspace{2 pt}
\includegraphics[width=0.24\textwidth]{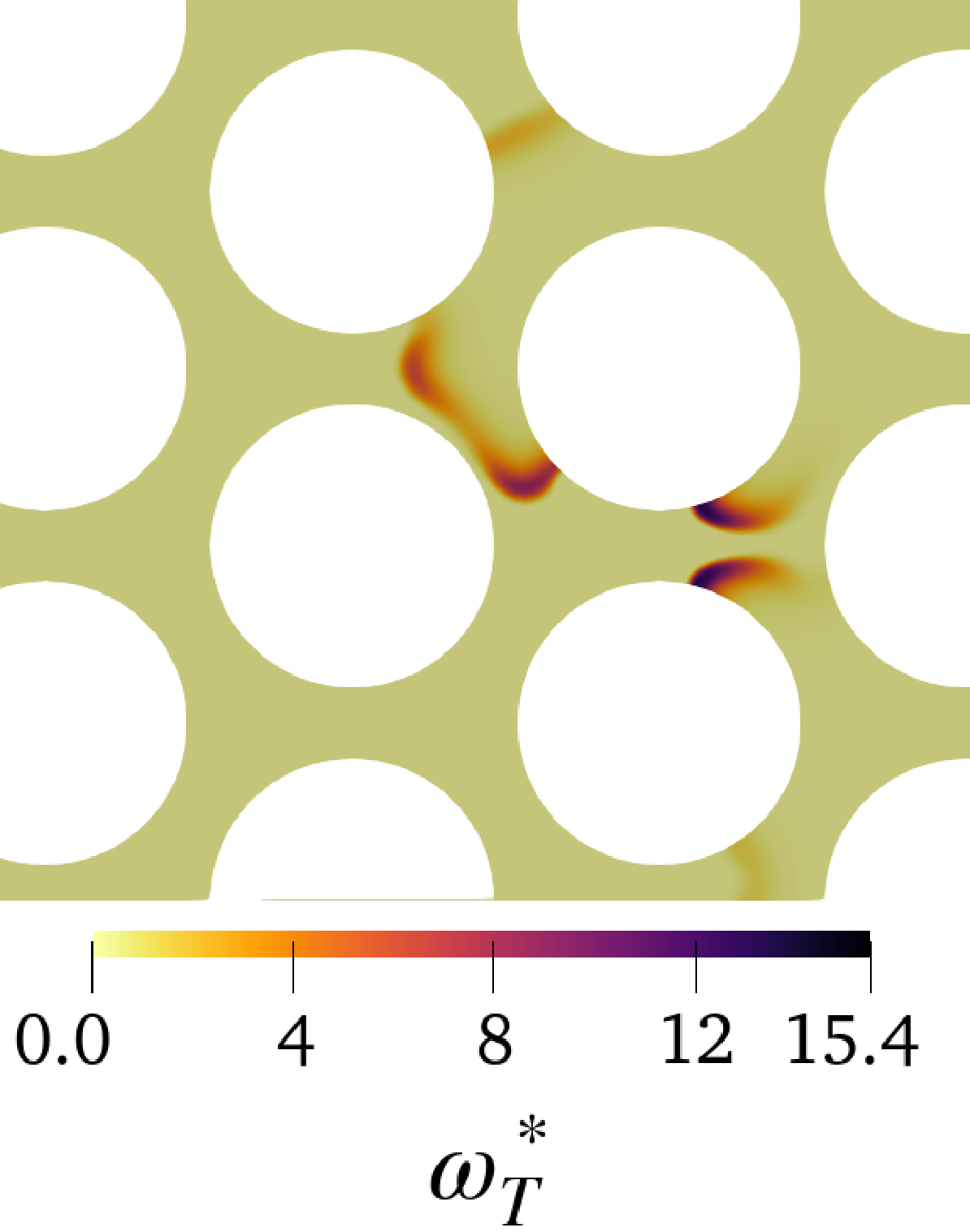}
\includegraphics[width=0.24\textwidth]{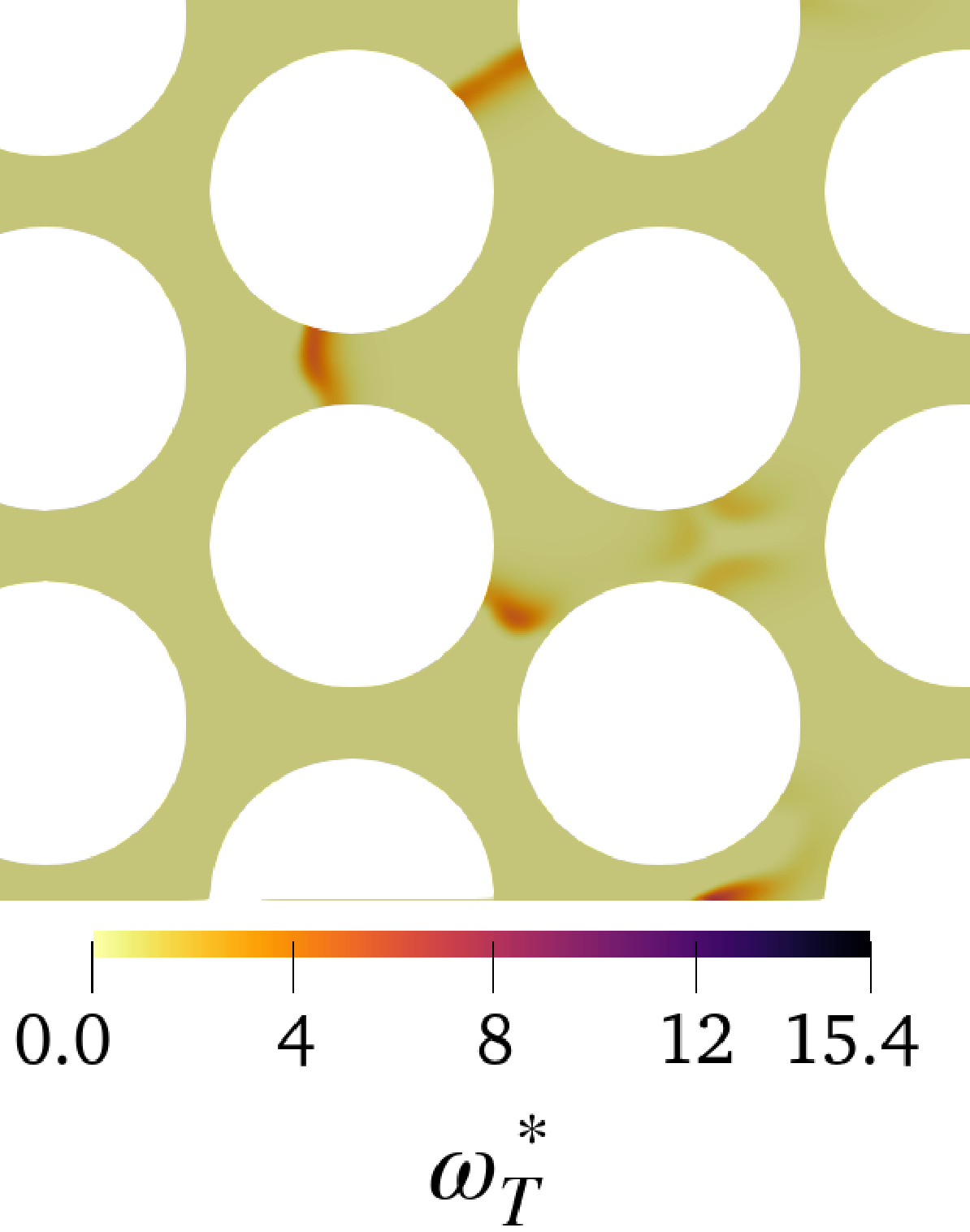}
\caption{Snapshots of heat release rate showing propagation of flame through typical pore with confinement $S/D=5/4$. From left to right: $t^*-t^*_0=0$, $t^*-t^*_0\approx0.42$, $t^*-t^*_0\approx0.77$, $t^*-t^*_0\approx0.98$.}
\label{fig:SD_54_pbp}
\end{figure*}

We now consider the effect of increasing confinement. Fig.\ref{fig:SD_54_pbp} displays the propagation of the flame through a typical pore in the matrix with confinement $S/D=5/4$. Here we observe somewhat different behaviour to the $S/D=5/2$ case. Focusing on the two complete pores in the second (from right) row of cylinders, although the flame is propagating approximately evenly at the initial $t^*=t^*_0$ time, as the flame evolves there is preferential propagation through the upper pore, though even in this single pore the flame propagates asymmetrically. In the third panel we observe that when the flame has propagated through the upper pore the flame in the lower pore has actually been pushed backwards slightly, and the maximum heat release rate is attained close to the cylinder surfaces. As the flame originating in the upper pore continues to enter into the lower pore it splits, with part of the flame propagating through another (diagonal) pore, and the rest burning backwards through the lower pore until the flame becomes locally extinct once the fuel in this region has been consumed. 

Fig.\ref{fig:flame_histograms} shows histograms of the local flame speed and thickness for the flame in Fig.\ref{fig:SD_54_pbp} across the time $0\leq t^*-t^*_0\leq0.98$. In contrast to the $S/D=5/2$ case, the PDF of the local flame thickness has two peaks. This behaviour is due to different propagation observed in each pore in Fig.\ref{fig:SD_54_pbp}. The flame in the lower pore, which does not propagate through the pore, retains a relatively constant thickness close the laminar value, however the flame front propagating through the upper pore becomes much thinner as the flame front splits, resulting in the peak at $\delta_{loc}/\delta_L= 0.5-0.6$. PDFs at each time-step showed significant variance, evidencing the rapidly changing flame properties in this confinement.
%\begin{figure*}[h!]
%\centering
%\vspace{0.4 i}
%\includegraphics[width=0.48\textwidth]{10x8y_2pore_fs.eps}
%\includegraphics[width=0.48\textwidth]{10x8y_2pore_ft.eps}
%\caption{PDF of normalised flame speed (left panel) and normalised flame thickness (right panel) for the part of the flame in the domain shown in Fig\ref{fig:SD_54_pbp} for time period $0\leq t^*-t^*_0\leq 0.98$.}
%\label{fig:flame_histograms_SD_54}
%\end{figure*}

\subsection{Large-Scale Flame Dynamics}
\label{sub_sec:large_scale_dynamics}
\begin{figure*}
    \centering
    \includegraphics[width=0.48\textwidth]{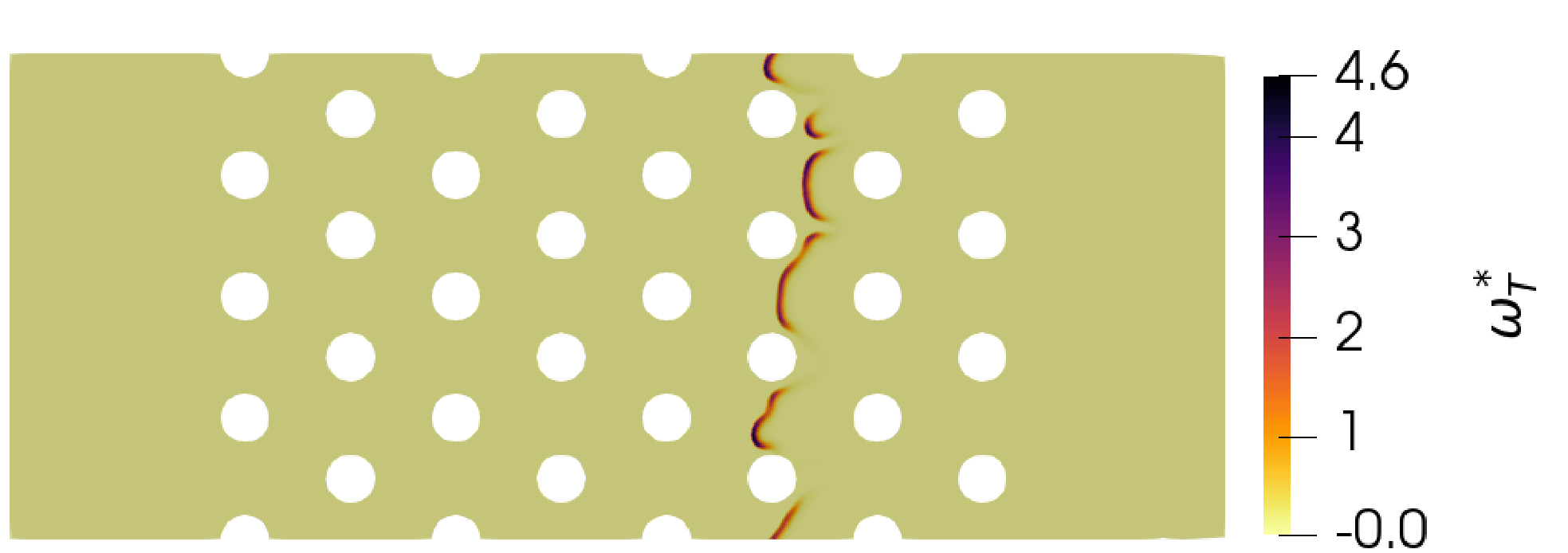}
    \includegraphics[width=0.48\textwidth]{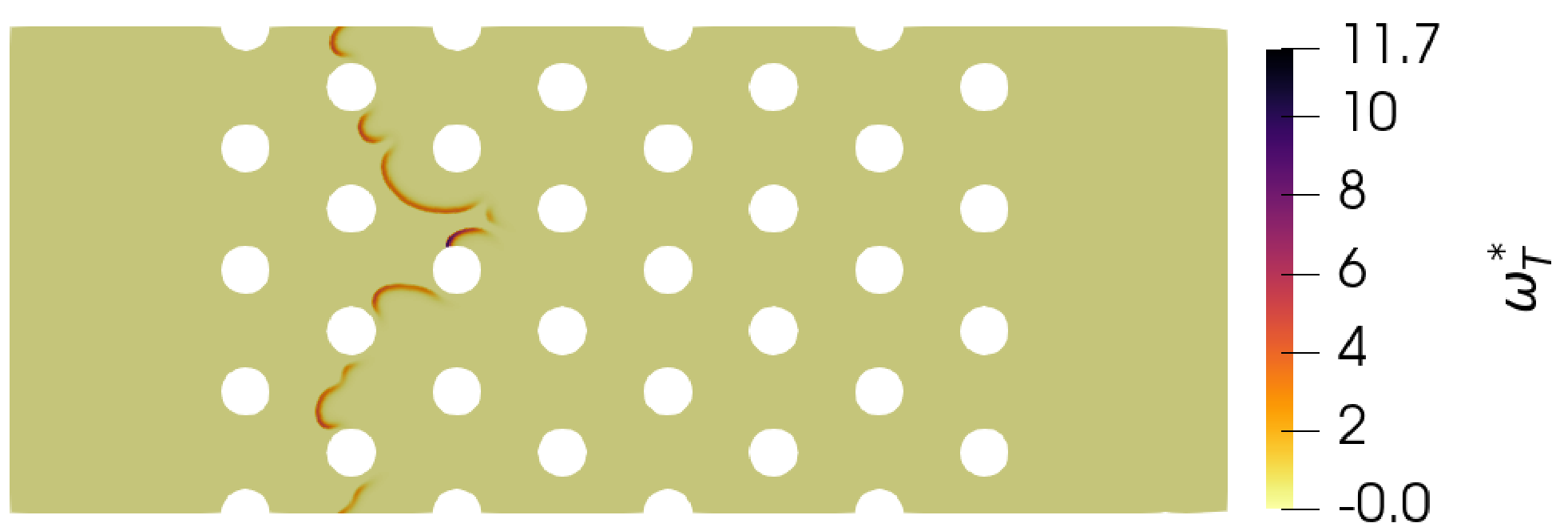}
    \caption{Isocontours of heat release rate displaying large scale flame behaviour for confinement $S/D=5/2$. Left panel $t^*\approx10.5$, right panel $t^*\approx20.0$.}
   \label{fig:hrr_snapshots_SD_52}
\end{figure*}

We now examine how confinement affects flame propagation on a larger scale. Fig.\ref{fig:hrr_snapshots_SD_52} shows snapshots of the heat release rate at two different time for confinement of $S/D=5/2$. At $t^*\approx 10.5$ the flame has propagated relatively evenly (in $x_2$) in the domain though some flame cusps are present. By $t^*\approx 20.0$, the `W' shapes associated with a thermodiffusive instability can be observed, along with greater super-adiabatic temperatures.
We can deduce from Fig.\ref{fig:hrr_snapshots_SD_52} that for this level of confinement, the structures characteristic of a thermodiffusive instability becomes fully developed on a time scale of $t^*=O(10)$. For comparison, in unconfined simulations we observe fully formed flame fingers and `W' shaped flame fronts by $t^*=8.0$. Confinement therefore suppresses the larger scale structures until later time.

\begin{figure*}
    \centering
    \includegraphics[width=0.48\textwidth]{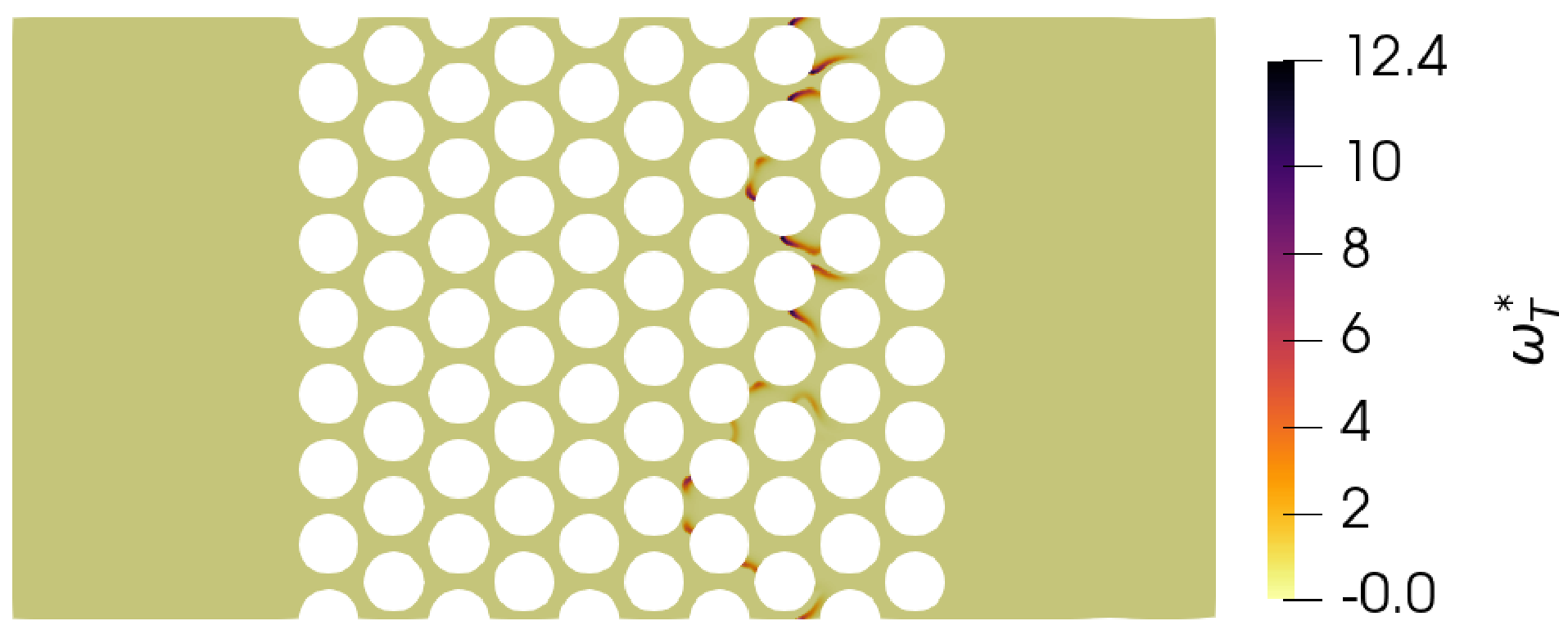}
    \includegraphics[width=0.48\textwidth]{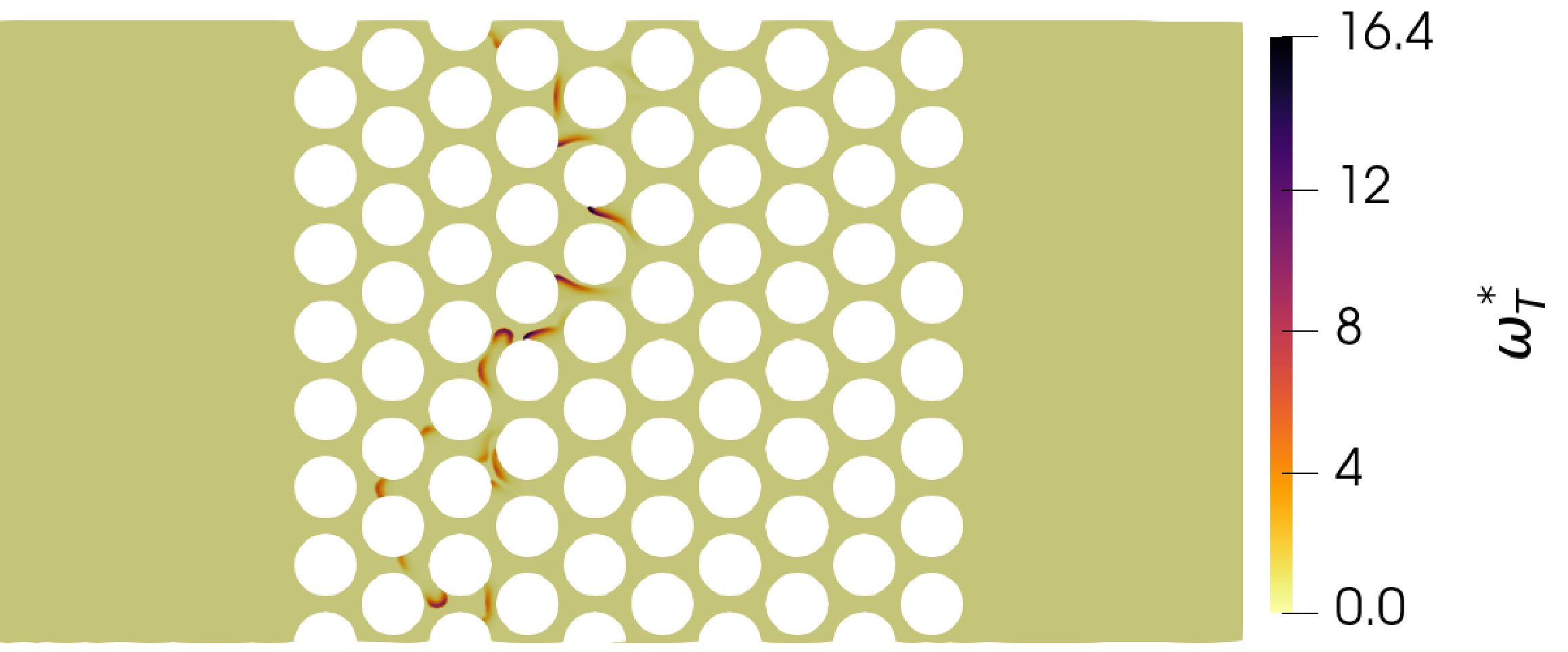}
    \caption{Isocontours of heat release rate displaying large scale flame behaviour for confinement $S/D=5/4$. Left panel $t^*\approx13.4$, right panel $t^*\approx14.1$.}
   \label{fig:hrr_snapshots_SD_54}
\end{figure*}

Fig.\ref{fig:hrr_snapshots_SD_54} likewise displays the heat release rate at two different times, though now the confinement is $S/D=5/4$. Even at early times we observe asymmetric propagation of the flame and large heat release rates. This behaviour continues as the flame evolves through the array, and no evidence of large scale thermodiffusive structures can be observed.

To explain the effect of confinement on flame propagation we heuristically apply the findings of flame stability theory. Once the flame is established in the array we see no evidence of the influence of a thermoacoustic instability, therefore at the equivalence ratio used (i.e. $\phi=0.32$) thermodiffusive effects are likely to be the dominant flame instability present. \cite{frankel_sivashinsky_1982}, \cite{pelce_clavin_1982}, \cite{matalon_matkowsky_1982} independently derived the dispersion relation describing the instability of the flame front, which can be expressed in the form (see~\cite{Matalon_2003})
\begin{equation}
    \omega=\omega_{DL} S_L k-\omega_{D} k^2.
\end{equation}
The first term corresponds to the Darrieus--Landau instability which is unconditionally unstable, and the second order term (which takes into account molecular diffusion, thermal diffusion, and viscous diffusion) can be stabilising or destabilising depending on the sign of $\omega_D$. For sufficiently small effective Lewis numbers this dispersion relation predicts a thermodiffusively unstable flame, with (moderately) small wavelength disturbances found to be more unstable than long wavelength modes (sufficiently small wavelengths are stabilised by a higher order term $O(k^4)$). In addition, the weakly nonlinear analysis of \cite{Sivashinsky_1977} (leading to the Michelson-Sivanshinsky equation) predicts that smaller wavelength modes coalesce into larger cells as the flame evolves, forming the `W' shape characteristic of thermodiffusive instabilities.

In our simulations, the throat width between cylinders limits the maximum initial wavelength of disturbance, with greater confinement reducing this maximum. Once the flame enters the array, local flame fronts are forced to split as the flame approaches individual cylinders. Flame cusps are prevented from coalescing into larger structures by this process, as flame fronts are continually broken up near boundaries. This creates short wavelength disturbances which have larger growth rates than larger scale structures, leading to quicker propagation through the array.

Increasing confinement both restricts the large scale structures that may be formed and forces flame fronts to split more regularly, causing a prevalence of short wavelength modes. Given these modes are associated with larger growth rates, the thermodiffusive mechanism predicts that increased confinement should result in a faster propagating, asymmetric flame, in agreement with our numerical results.

\subsection{Further Increasing Confinement}
\label{sub_sec:TA_instabilities}
\begin{figure*}
\centering
\vspace{-1pt}
\includegraphics[width=0.24\textwidth]{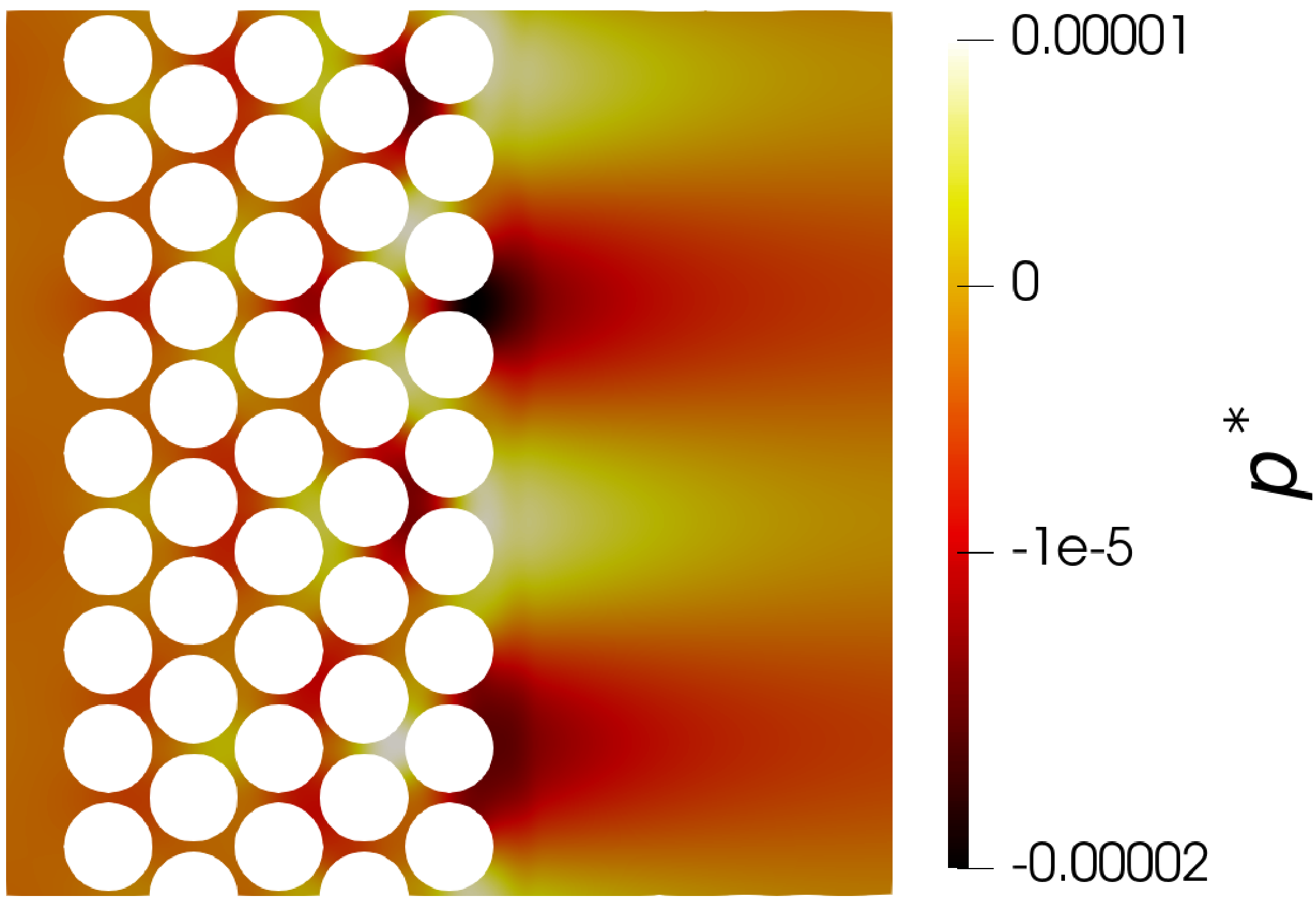}
\includegraphics[width=0.24\textwidth]{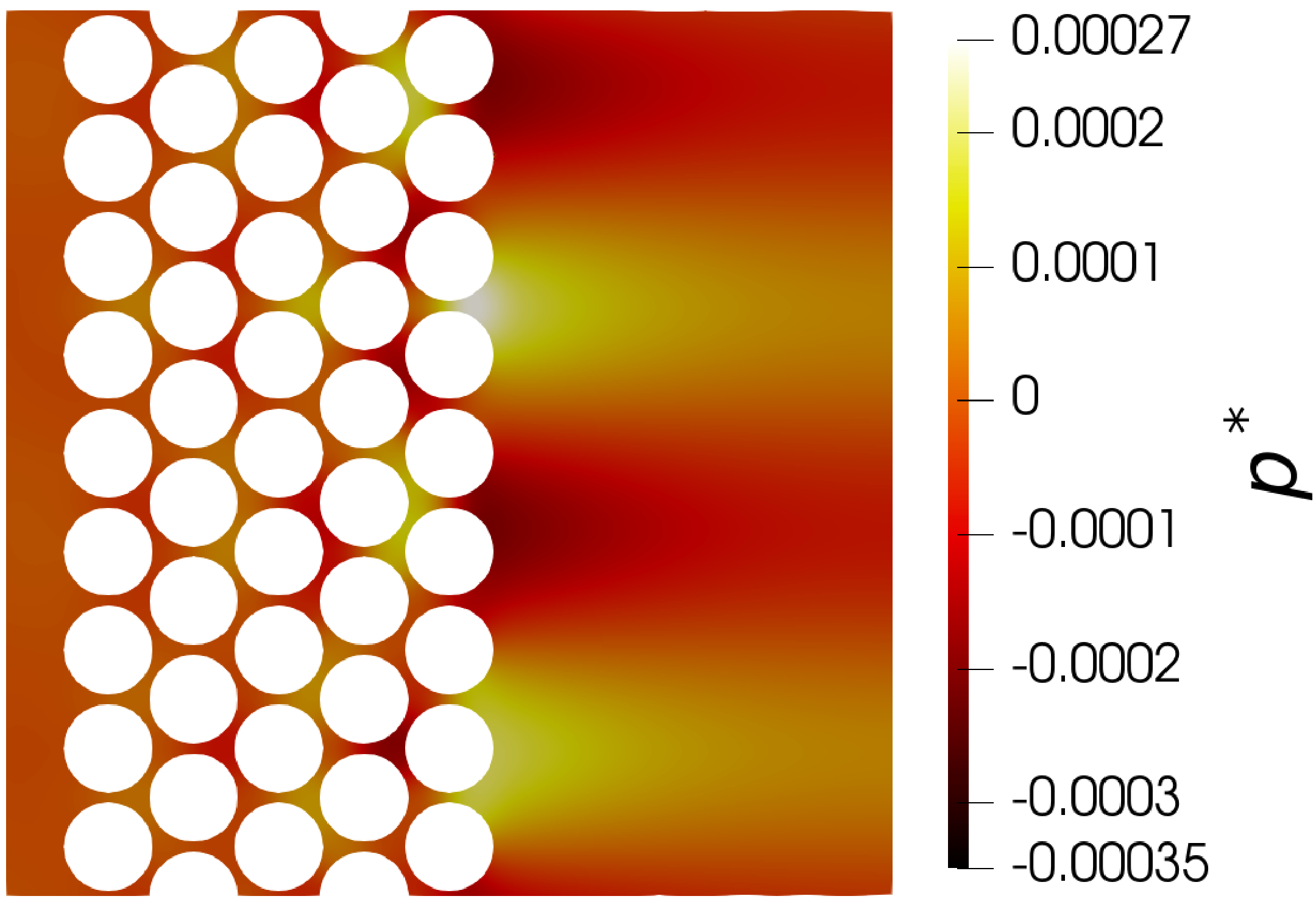}
%\vspace{2 pt}
\includegraphics[width=0.24\textwidth]{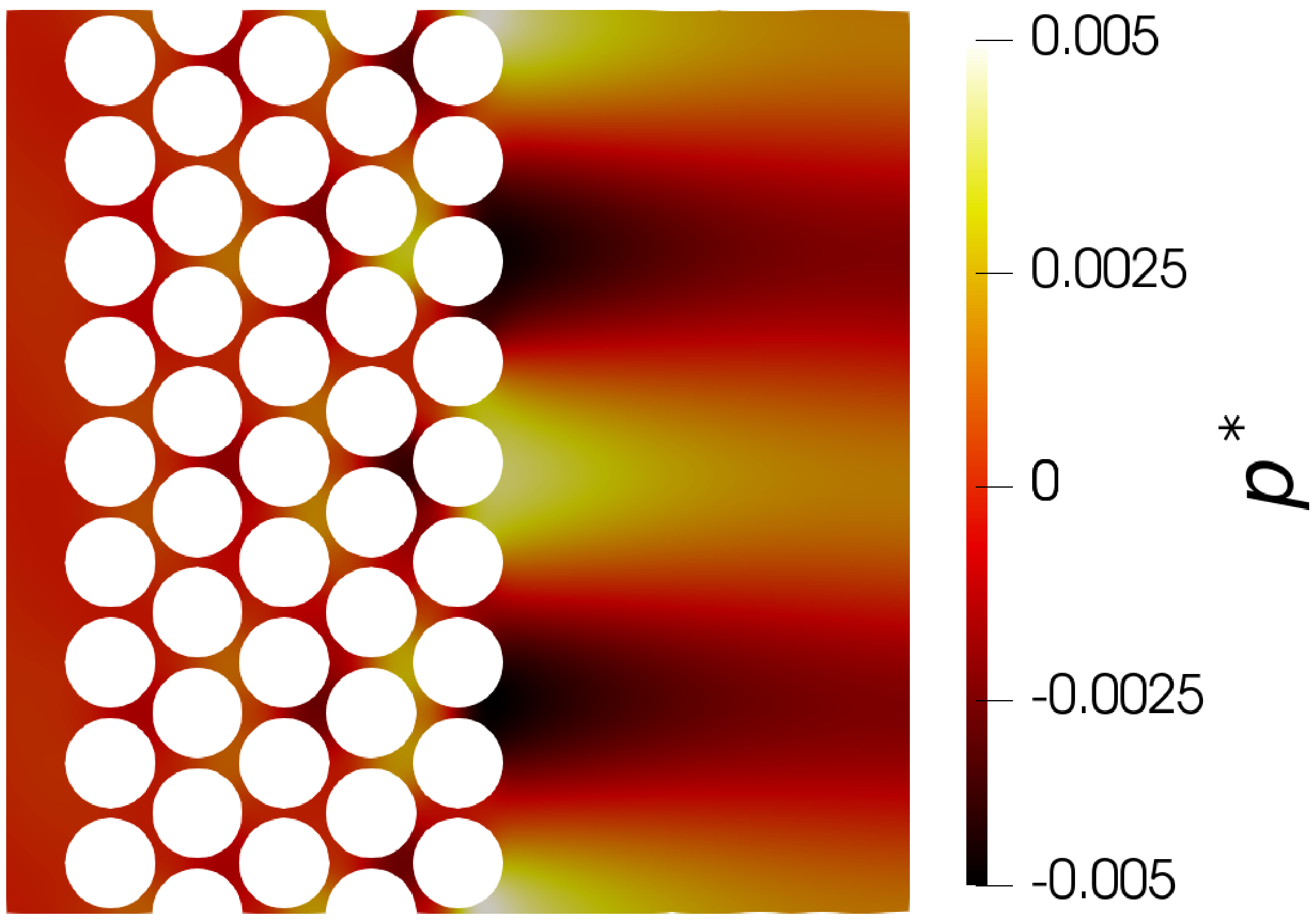}
\includegraphics[width=0.24\textwidth]{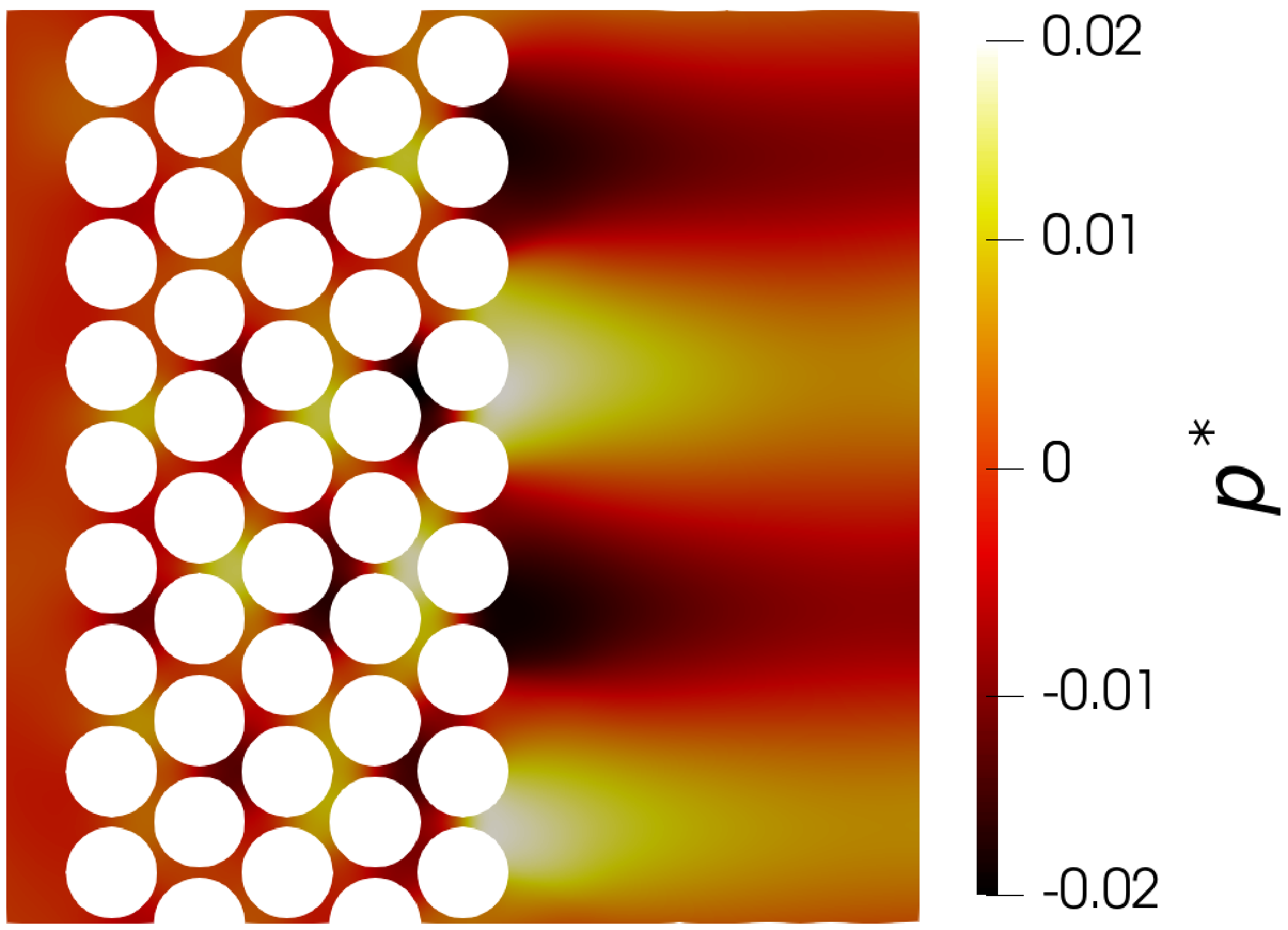}
\caption{Snapshots of pressure for confinement of $S/D=10/9$ at increasing times. From left to right $t^*\approx 0.63$, $t^*\approx 1.33$, $t^*\approx 2.03$, $t^*\approx 2.38$. }
\label{fig:SD_109_pressure}
\end{figure*}

We next consider the case with confinement $S/D=10/9$. Fig.\ref{fig:SD_109_pressure} shows the pressure field at various times as the flame approaches the array. A subharmonic travelling (in $x_2$) pressure wave lies within the first three rows of the porous matrix, and as the flame is drawn towards the array this wave grows rapidly in amplitude. As a crude estimate of the growth rate of this instability we assumed that the maximum amplitude (in the entire domain), of the pressure perturbations grew like $e^{\sigma^* t^*}$ at early times, giving a growth rate $\sigma^*$ of
\begin{equation}
\label{eq::TA_growth_rate}
    \sigma^*=3.9\pm 0.2
\end{equation}
for $0.63<t^*<2.38$. Eventually, these oscillations could not be resolved and the numerical solution diverged, with identical behaviour observed at a similar time when the resolution was doubled. This behaviour is typical of a thermoacoustic instability which are known to present major obstacles in the design of durable, efficient, hydrogen combustors. We note that when the growth rate given in \eqref{eq::TA_growth_rate} is made dimensional it is of the same order of magnitude as those found in previous experimental studies of hydrogen flames in other contexts (see for example \cite{Vaysse_2024}). Travelling pressure waves were also observed prior to the flame entering the array in the less confined cases present above, though they quickly dissipated once the flame began to propagate through porous matrix.

The presence of the thermoacoustic instability observed here demonstrates that care is required when simulating hydrogen flames propagating through porous media. Indeed, when the single-step reaction mechanism of \cite{dominguez_2023} was used, thermoacoustic instabilities were observed to grow much more rapidly and often led to divergence of the numerical solution.

\section{Conclusions}\label{sec:conc}
We have investigated the dynamics of lean hydrogen flames propagating through arrays of cylinders using a high-order mesh-free framework for direct numerical simulations. In this setting, beyond the capabilities of leading high-order mesh-based combustion codes, thermodiffusive and thermoacoustic instabilities play a role. This work demonstrates the potential of the mesh-free approach as an alternative to traditional mesh-based combustion DNS.

Increased confinement leads to, on average, faster, thinner, hotter flames. On a pore scale greater confinement leads to uneven propagation even through the first few rows of pores, with a larger variance in local flame speed and thickness. Increasing confinement limits the maximal initial wavelength of perturbation to the flame front in each pore, preventing the formation of large scale structures commonly associated with thermodiffusive instabilities. Given that shorter wavelength modes grow quicker than their larger wavelength counterparts, greater confinement led to quicker growth of the thermodiffusive instabilities and a large variance in propagation on a pore-by-pore basis.

At sufficient confinement levels, flames become susceptible to thermoacoustic instabilities prior to entry of the porous array. Whilst these instabilities quickly dissipate if the flame enters the porous array, for the most confined cases the instabilities grow exponentially. Practical combustor designs must therefore weigh the potential benefits of using confinement against damage to combustors by large amplitude thermoacoustic waves. Studies on three-dimensional effects on flame instabilities in canonical porous geometries are ongoing.

\section{Declaration of competing interest} \addvspace{2pt}

The authors declare that they have no known competing financial interests or personal relationships that could have appeared to influence the work reported in this paper.

\section{Acknowledgments} \addvspace{2pt}

This work was partially funded by the Engineering and Physical Sciences Research Council (EPSRC) grant EP/W005247/2. JK is supported by the Royal Society via a University Research Fellowship (URF\textbackslash R1\textbackslash 221290). We thank EPSRC for computational time made available on the UK supercomputing facility ARCHER2 via the UK Turbulence Consortium (EP/X035484/1). We are grateful for assistance given by Research IT and the use of the Computational Shared Facility at the University of Manchester. 

\bibliography{apssamp}% Produces the bibliography via BibTeX.

\end{document}